\documentclass{statsoc}

\usepackage{geometry}

\usepackage[utf8]{inputenc}

\usepackage{amssymb,stmaryrd}
\usepackage{amsmath,amsbsy,natbib,amsfonts}
\usepackage[colorlinks,citecolor=red,urlcolor=red]{hyperref}
\usepackage{graphicx,color} % support the \includegraphics command and options
\usepackage{arydshln}
\usepackage{dashrule}
\usepackage{nccmath,lipsum}

\newcommand{\pip}{\pi^\pr}

\newcommand{\vecpi}{\tilde{\pi}}
\newcommand{\vecpip}{\tilde{\pi}^\pr}

\newcommand{\vectau}{\tilde{\tau}}

\newcommand{\vecx}{\tilde{x}}
\newcommand{\vecy}{\tilde{y}}
\newcommand{\vecv}{\tilde{v}}

\newcommand{\vecHp}{\tilde{H}^\pr}

\newcommand{\hatpi}{\hat{\pi}}
\newcommand{\hattau}{\hat{\tau}}

\newcommand{\dash}{\mbox{-}}

\newcommand{\eff}{\xi}
\newcommand{\rrp}{\lambda}

\newcommand{\diag}[1]{D_{#1}}

\def\ng{%
  \setbox0=\hbox{-}%
  \vcenter{%
    \hrule width\wd0 height \the\fontdimen8\textfont3%
  }%
}

\newcommand{\hatr}{{\hat r}}

\newcommand{\haty}{{\hat y}}

\newcommand{\real}{{\mathbb{R}}}

\newcommand{\hattheta}{{\hat{\theta}}}

\newcommand{\hatrho}{{\hat{\rho}}}

\newcommand{\bcb}{\begin{color}{blue}}
\newcommand{\bcr}{\begin{color}{red}}
\newcommand{\bcg}{\begin{color}{green}}
\newcommand{\ec}{\end{color}}

\newcommand{\bc}{\begin{color}{green}}

\newcommand{\pr}{{\prime}}

\newcommand{\tx}{{\tilde{x}}}

\newcommand{\ty}{\tilde{y}}
\newcommand{\tg}{\tilde{g}}
\newcommand{\tbeta}{\tilde{\beta}}
\newcommand{\htheta}{\hat{\theta}}

\newcommand{\var}[1]{{\mbox{Var}}[#1]}

\newcommand{\corr}[2]{{\mbox{Cor}}\left[#1, #2\right]}

\newcommand{\vars}[2]{\mbox{Var}_{#2}[#1]}

\newcommand{\bdm}{\begin{displaymath}}
\newcommand{\edm}{\end{displaymath}}

\newcommand{\beq}{\begin{equation}}
\newcommand{\eeq}{\end{equation}}

\newcommand{\bea}{\begin{eqnarray}}
\newcommand{\eea}{\end{eqnarray}}

\newcommand{\beas}{\begin{eqnarray*}}
\newcommand{\eeas}{\end{eqnarray*}}

\newcommand{\bdf}{\begin{defn}}
\newcommand{\edf}{\qed\end{defn}}

\newcommand{\bex}{\begin{ex}}
\newcommand{\eex}{\qed\end{ex}}

\newcommand{\bthm}{\begin{theorem}}
\newcommand{\ethm}{\qed\end{theorem}}

\newcommand{\argmin}{\mbox{argmin}}
\newcommand{\argmax}{\mbox{argmax}}

\newcommand{\bsp}{\begin{sloppypar}}
\newcommand{\esp}{\end{sloppypar}}

\newcommand{\ttt}[1]{\texttt{#1}}

\newcommand{\mi}{\text{-}}

\newcommand{\pd}[2]{\frac{\partial #1}{\partial #2}}

\newcommand{\hbeta}{\hat{\beta}}

\newcommand{\bcx}{\begin{color}{blue}}

\title[Compound Logistic Regression]{A Compound Logistic Regression Model for Binary Responses}

\author[A~Almudevar]{Anthony Almudevar, PhD}
\address{Department of Biostatistics and Computational Biology, University of Rochester, 
Rochester, NY,
USA.}
\email{anthony\_almudevar@urmc.rochester.edu}
\author[J~Almudevar]{Jacob Almudevar, MSc}
\address{Department of Mathematics and Statistics, University of New Hampshire, 
Durham, NH,
USA.}

\begin{document}

\begin{abstract}
Logistic regression is the most commonly used method for constructing predictive models for binary responses. One significant drawback to this approach, however, is that the asymptotes of the logistic response function 
are fixed at 0 and 1, and there are many applications for which this constraint is inappropriate.  More flexible models have been proposed for this application, most proceeding by supplementing the logistic response function with additional parameters. In this article we extend these models to allow correlated responses and the inclusion of covariates. This is achieved through the \emph{compound logistic regression model}, for which the mean response is a  function of several logistic regression functions. This permits a greater variety of models, while retaining the advantages of logistic regression.  \\

%We applied the method using longitudinal data for seven antibodies for candidate Streptococcus pneumonia antigens assessed for COP against acute otitis media (AOM) infection in children,  including as a covariate otitis %proneness (OP) (occurrence of 3 cases of AOM within 6 months or 4 cases in a year). The model allowed decomposition of the COP property into protective threshold (minimum protective antibody concentration); exposure to i%nfection risk; and vaccine efficacy. The effect of  OP on each component could be assessed separately, and was found to affect vaccine efficacy, supporting the hypothesis of OP as a true immunological deficiency.   \\

The methods are implemented in an \ttt{R} package \ttt{CLmodel} (\url{https://github.com/ALMUDEVAR163/CLmodel/}).

%An immunological correlate of protection (COP) is a biomarker, typically a measure of antibody  (Ab) concentration, which is negatively associated with infection risk. The COP property can be expressed mathematically as a function which maps biomarker levels to infection  risk. This function can be estimated from paired biomarker/infection event data. Although the logistic regression model is commonly used to model binary responses, we found that it is not sufficiently general for this application, since asymptotes other than 0 or 1 cannot be anticipated.  On the other hand, logistic regression, as a generalized linear model, offers considerable advantages, including the natural inclusion of covariates and the modelling of response correlation. 

%In this article we show how such modelling extensions can be implemented for a double-asymptotic model, in which the asymptotes may be fully modeled.  The proposed fitting methods are applied to a set of simulated models, and to a longitudinal data set in which assays of seven antibodies for candidate  \textit{Streptococcus pneumonia} antigens are paired with acute otitis media (AOM) infection events in young children. We found that the various features of COP as a function of antibody concentrations can be reliably estimated with the inclusion of covariates and response correlation models. The COP property is shown to be statistically significant for 5 of the 7 candidates, and evidence of an age effect modifying the relationship between antibody concentration and infection risk is reported. 

\end{abstract}


%\maketitle

%\footnotetext[2]{Please ensure that you use the most up to date
%class file,
%available from the SIM Home Page at\\
%\href{http://www.interscience.wiley.com/jpages/0277-6715}{\texttt{www.interscience.wiley.com/jpages/0277-6715}}}

\vspace{6cm}

\section{Introduction}\label{sec.introduction}

%REDO INTRO We are given paired observations $(x_i,y_i)$, $i = 1,\ldots,n$, where $x_i$ is a $p$-dimensional vector of covariates and $y_i$ is a binary response. We will eventually admit the possible that the responses are statistically dependent.  \vspace{2cm}

We are given paired observations $(x_i,y_i)$, $i = 1,\ldots,n$, where $x_i$ is a $p$-dimensional vector of covariates and $y_i$ is a binary response, interpreted as a Bernoulli random variable.  The problem is to use the observations to estimate a model of the form 
\beq
P(y = 1\mid x  ) = \pi(x), \,\, y \in \{0,1\},\,\, x \in \real^p.  \label{eq.binary.model}
\eeq
To motivate the methodology proposed in this article we will initially assume that $x = z \in \real^1$ is some biomarker able to predict  the presence or absence of  a condition,  coded as $y = 1$ or $y=0$. We might reasonably expect response function $\pi(z)$ to be monotone in $z$ and bounded in $[0,1]$, so there must exist asymptotes:
$$
a_L = \lim_{z \rightarrow\,\mi\infty} \pi(z) \,\,\, \mbox{and}\,\,\,  a_U = \lim_{z \rightarrow\,\infty} \pi(z).
$$
If $\pi(z)$ is increasing, then $a_U$ ($a_L$) is the prevalence of the condition among subjects with high (low) levels of the biomarker.  One problem of importance is the identification of a \emph{threshold}  level $z_{thr}$, that is, a value near which $\pi(z)$  transitions rapidly between the two asymptotes. If $\pi(z)$ possesses odd symmetry, then $z_{thr}$ can be defined as the point of symmetry, otherwise some canonical definition of $z_{thr}$ can be used.  We might also introduce a fourth quantity $\alpha$ to represent the rate of increase of $\pi(z)$ in the neighborhood of $z_{thr}$. %%%We will review in this section a class of models permitting a precise parametric definition of $(a_1, a_2, z_{thr}, \alpha)$. 

To fix ideas, suppose $z$ is some antibody concentration and $y$ codes an infection event.  A pure \emph{threshold model} is proposed in  \cite{chen2013}, given as
\beq
\pi_{inf}(z) = a_L I\{z < z_{thr} \} + a_U I\{z \geq z_{thr}\}, \label{eq.chen.threshold}
\eeq  
where $z_{thr}$ is a "correlate of protection" (COP), that is, a concentration of antibody sufficient to confer some level of immunity. Then $a_U$ and $a_L$ are the infection rates with and without that protection, respectively. Here,  $\pi_{inf}(z)$ is simply a discrete step function, with $\alpha$ playing no role (equivalently, $\alpha = \infty$).  In contrast, in a study reported in   \cite{chan2002} protection was observed to vary continuously with antibody levels. A general COP model should anticipate both cases.  %Thus, a distinction in the literature is made between cases for which protection is dichotomous, with infection risk satisfying $\pi(z) \approx a_2$ or $\pi(z) \approx a_1$, for $z > z_{thr}$ or $z < z_{thr}$, or for which protection is continuously dependent on $z$. The parameter $\alpha$ will serve as a continuous index for the latter case.  

Undoubtedly the most commonly used response function for binary response models is the logistic function $\phi(u) = 1/(1 + \exp(\mi u))$. Incorporating a single covariate with an intercept term, the logistic  regression model is commonly written
\beq
\pi(z) = \phi(\beta_0 + \beta_1 z). \label{eq.conventional.logistic.regression}
\eeq  
With a simple one-to-one reparametrization  Equation \eqref{eq.conventional.logistic.regression} can be written
\beq
\pi(z) = \phi(\alpha (z - z_{thr})) \label{eq.conventional.logistic.regression.cop}
\eeq  
to match the elements of the COP model. However, this model is constrained to have asymptotes equal to zero or one, which would be unreasonably restrictive for many applications of this type. Of course, it is possible to construct a generalized logistic function by incorporating additional parameters, for example,    
\beq
\phi_g(t) = a + \frac{b-a}{(c + e^{- d  -  f t})^{1/g}} \label{eq.generalized.logistic.function}
\eeq
for suitably bound constants $(a,b,c,d,f,g)$ \citep{richards1959}. 
%\subsection{Binary COP model}\label{sec.binary.cop.model}

We next argue that a reasonable level of generalization  can be achieved by setting $c = g =1$ in Equation \eqref{eq.generalized.logistic.function}, which is able to extend model \eqref{eq.conventional.logistic.regression.cop} to
\beq
\pi_g(z) = a_L + \frac{a_U - a_L}{1 + e^{\mi \alpha(z - z_{thr})}}, \label{eq.general.cop}
\eeq
using the parameters $(a_L,a_U, z_{thr}, \alpha)$  introduced  above. Of course, the advantage of the logistic model of Equation \eqref{eq.conventional.logistic.regression} is that $z$ can be replaced with a vector of covariates $\tx \in \real^p$, with coefficient vector $\tbeta \in \real^p$, the model now becoming  $\pi(\tx) = \phi(\eta)$, $\eta = \tbeta^T\tx$ (note that throughout this article, in the context of matrix algebra any vector is assumed to be a column vector). This allows the response distribution to depend on multiple covariates in an intuitive and analytically tractable way. 

The purpose of this article is to propose a method of applying this kind of multivariate extension to a model such as Equation \eqref{eq.general.cop}. To this effect we may always rewrite \eqref{eq.general.cop} as
\beq
\pi_g(\tx) = a_L + \frac{a_U - a_L}{1 + e^{\mi \eta}} = a_L + (a_U - a_L)\phi(\eta), \label{eq.general.cop.eta}
\eeq
making use of the linear term $\eta = \tbeta^T\tx$. Then through a simple reparametrization  
$\eta$ can include  $\alpha(z - z_{thr})$,  as well as additional covariates on which the effective COP threshold would be allowed to depend. 

However, we would also like to allow the asymptotes $a_L$, $a_U$ to depend on additional covariates, involving coefficients that admit conventional forms of inference and intuitive interpretations. We next introduce a method by which this may be done.  

\subsection{Asymptote Model}

Following Equation \eqref{eq.general.cop.eta}, since $a_1$ and $a_2$ are probabilities, we can reasonably represent each by its own logistic response function. Thus, Equation \eqref{eq.general.cop.eta} can be revised to give
\bea
\pi_c(\tx_1, \tx_2, \tx_3) &=& \phi(\eta_2) +  (\phi(\eta_3) - \phi(\eta_2)) \phi(\eta_1)  \nonumber \\
&=&  \phi(\eta_3) \phi(\eta_1)  +  \phi(\eta_2) (1 -  \phi(\eta_1))  \label{eq.general.cop.etac}
\eea
where $a_L = \phi(\eta_2)$ and  $a_U = \phi(\eta_3)$, and $\eta_j = \tbeta_j^T \tx_j$, $j = 1,2,3$. We assume $\tx_1$ incorporates $z$, and any other covariates predictive of $z_{thr}$, while $\tx_2$ and $\tx_3$ incorporate covariates predictive  of $a_L$ and $a_U$ respectively. 

The relationship between the logistic response  components in Equation \eqref{eq.general.cop.etac} can be summarized using a single compounding function $H$, for example,
\bea
\pi_c(\tx_1, \tx_2, \tx_3) &=& H(\phi(\eta_1), \phi(\eta_2),  \phi(\eta_3)),† \mbox{ where } \nonumber \\ 
H(u_1, u_2, u_3) &=&  u_3u_1 + u_2(1 - u_1). \label{eq.general.cop.etac.H}
\eea

\subsection{Vaccine Efficacy Model}

We may, however, consider alternative compounding methods.  For example, compare model \eqref{eq.general.cop} to a COP model considered in the literature: 
\beq
\pi_{inf}(z) = \eff  (1 - \gamma \kappa(z)), \label{eq.dunning.scaled.logit}
\eeq
\citep{dunning2006,dunning2015}.  Here, $\kappa(z)$ models COP threshold, and is therefore a function of antibody $z$, with lower and upper asymptotes 0 and 1, representing minimum and maximum protection. This can be reasonably modeled by 
\beq
\kappa(z) = \phi(\alpha(z - z_{thr})), \label{eq.dunning.scaled.logit.b}
\eeq
in which case \eqref{eq.dunning.scaled.logit} and \eqref{eq.dunning.scaled.logit.b} are together equivalent to \eqref{eq.general.cop}. In particular,  substituting $\kappa(z) = 0,1$ gives unprotected risk $\pi_{inf}^{max} = \eff = a_L$, and fully protected risk $\pi_{inf}^{min} = \eff(1 - \gamma) = a_U$. Note that the threshold  model \eqref{eq.chen.threshold} is approached in the limit as $\alpha \rightarrow\infty$.  Thus, $\eff$ can be interpreted as infection risk attributable to exposure. In addition, recall the definition of  vaccine efficacy
$$
VE = 1- \frac{r_v}{r_e}
$$
where $r_v$ is the rate of disease among vaccinated individuals and $r_e$ is the rate among unvaccinated individuals. Here, we may set $r_e = \eff$, $r_v = \eff(1 - \gamma)$, therefore
$$
VE = 1- \frac{\eff(1-\gamma)}{\eff} = 1 - (1 - \gamma) = \gamma \in (0,1),
$$
so that $\gamma$ is directly interpretable as vaccine efficacy. 

We note at this point that the threshold, efficacy and exposure components $\kappa(z), \gamma$ and $\eff$ capture quite distinct contributions to $\pi_{inf}(z)$. This must be anticipated by any methodology attempting to estimate the effect of covariates on a COP model. For example, one covariate may be expected to affect only exposure, another only vaccine efficacy, a third only COP threshold $z_{thr}$. In this case we may  substitute separate logistic response functions  $\kappa(z) = \phi(\eta_1)$, $\gamma = \phi(\eta_2)$ and $\eff = \phi(\eta_3)$, yielding 
\beq
\pi_{inf}(\tx_1, \tx_2, \tx_3) = \phi(\eta_3) (1 -  \phi(\eta_2) \phi(\eta_1))  \label{eq.general.cop.etac.2}
\eeq
(note that while $\gamma$ is not a probability we still expect it to be in the interval $(0,1)$).  As in Equation \eqref{eq.general.cop.etac.H} we can isolate the compounding function $H$, 
\bea
\pi_{inf}(\tx_1, \tx_2, \tx_3) &=& H(\phi(\eta_1), \phi(\eta_2),  \phi(\eta_3)), \mbox{ where } \nonumber \\ 
H(u_1, u_2, u_3) &=&  u_3(1 - u_2u_1). \label{eq.general.cop.etac.H2}
\eea
Note that the compounding function $H$ used in Equation \eqref{eq.general.cop.etac.H2} differs from that used in Equation \eqref{eq.general.cop.etac.H}.   We will show below that a considerable advantage of the compound logistic model is that the contributions to model complexity due to the multiple components scale in a linear manner, in the sense that the estimating equations for the model are a weighted sum of estimating equations constructed for each component separately (see Equation \eqref{eq.compound.model.2} below).  

\subsection{Interpretation of Compound Logistic Models}

At this point we have proposed constructing new response functions for binary response models by compounding  separate logistic 
response functions into a single expression defined by some compounding function $H$.  Distinct examples are given in 
Equations \eqref{eq.general.cop.etac.H}  and \eqref{eq.general.cop.etac.H2}.  We have not imposed any formal restrictions, other than 
noting that any logistic component $\phi(\eta_j)$ is, so far, being used to represent a probability, or some other quantity confined to the interval $(0,1)$. 

The natural interpretability of the logistic regression model $\pi(\tx) = \phi(\tbeta^T\tx)$ is well known.  For example, $\tbeta^T\tx$ is the predicted log-odds of $P(y=1)$ for an observation with covariates $\tx$, alternatively, $\beta_j \Delta$ is the predicted log-odds ratio between two observations which differ only in covariate $x_j$ by $\Delta$.  

We next show how similar interpretive statements  hold for the compound logistic models considered here. To do so, 
we assume the existence of some biomarker $z$,  and a relationship to a binary outcome $y$ governed by parameters
 $(a_1,a_2, z_{thr}, \alpha)$ as described above.  

\subsubsection{Asymptote model}\label{sec.asymptote.model}

Suppose, following  Equation \eqref{eq.general.cop.etac} we write
$$
\pi_c(\tx_1, \tx_2, \tx_3) = \phi(\eta_3) \phi(\eta_1)  +  \phi(\eta_2) (1 -  \phi(\eta_1)). 
$$
The biomarker $z$ is assumed to exist only in $\eta_1$, so that $z$ represents a natural axis for the model. Suppose interest is in comparing observations with high and low values of $z$. We can, without loss of generality, assume $\phi(\eta_1)$ is increasing in $z$ (otherwise, reverse the sign of the appropriate coefficient).  Then replace  $\phi(\eta_1)$ in \eqref{eq.general.cop.etac} with a general variable $U$, so that  
$$
\pi_c(U, \tx_2, \tx_3) = \phi(\eta_3) U  +  \phi(\eta_2) (1 -  U). 
$$
Then $U =  1$ is the limiting case as $z \rightarrow \infty$ and  $U = 0$ is the limiting case as $z \rightarrow -\infty$, interpretable simply as high or low biomarker levels, respectively (or maximum or  minimum biomarker effect). Then $P(y = 1 \mid U = 1) =  \phi(\eta_3)$ and  $P(y = 1 \mid U = 0) =  \phi(\eta_2)$, leading to the odds ratio
$$
\frac{Odds(y = 1 \mid U = 1)}{Odds(y=1  \mid U=0)} = e^{\eta_3 - \eta_2}.
$$
Thus, if we wish to allow this odds ratio to depend on a specific covariate,  we would need to include it in both $\tx_2$ and $\tx_3$. In Section \ref{sec.ex.asymptote.model} below we will consider 
the implications of doing so with respect to model identifiability and the stability of numerical fitting algorithms.

\subsubsection{Vaccine efficacy model}\label{sec.vaccine.efficacy.model}

\begin{figure}[h]
\centering
\includegraphics[height=2.25in, width=4.0in,viewport = 120 160 630 445, clip]{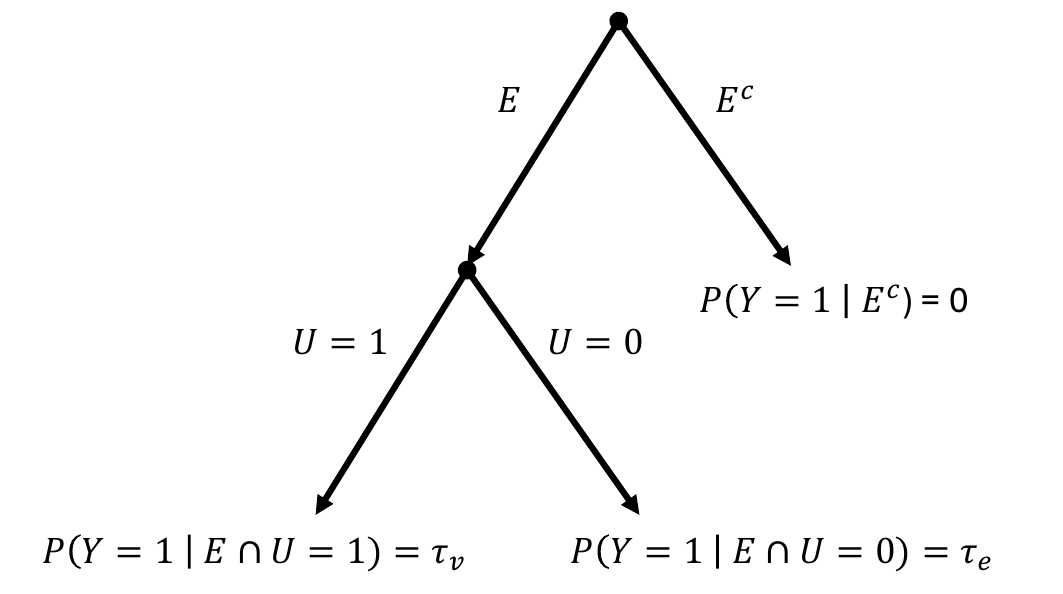}
\caption{Decision tree representation of infection/COP process introduced in Section \ref{sec.vaccine.efficacy.model}.}\label{ExposureModelFig01}
\end{figure}

Consider the sequence of events represented by the decision tree in Figure \ref{ExposureModelFig01}. For any individual, event $E$ is exposure to a pathogen. We assume an individual is either maximally protected (event $U = 1$) or has no protection (event $U = 0$). Then $y = 1$ if an individual  is infected or acquires an infectious disease (whatever the COP is intended to prevent), and is zero
otherwise. We then have the conditional probabilities
\beas
P(y = 1 \mid E^c) &=& 0, \\
P(y = 1 \mid E \cap U = 1) &=& \tau_v, \\
P(y = 1 \mid E \cap U = 0) &=& \tau_e.
\eeas
We also assume the events $E, \{U = 1\}$ are independent. We can observe $y$ and $U$ for an individual. Then
\bea
P(y=1 \mid U=1) &=& P(y=1 \mid E \cap U=1) P(E \mid U=1) \nonumber \\
&& \,\,\, + P(y=1 \mid E^c \cap U = 1) P(E^c \mid U = 1) \nonumber \\   
 &=& \tau_v P(E), \label{eq.exposure.model.1}
\eea
and similarly 
\bea
P(y=1 \mid U = 0) &=& \tau_e  P(E), \label{eq.exposure.model.2}
\eea
Suppose we set vaccine efficacy 
$$
\gamma = 1 - \frac{P(y=1 \mid U = 1)}{P(y=1 \mid U = 0)}.
$$  
The infection rate as a function of protection status can be expressed as
\beq
\pi_{E,U}(u) = P(y = 1 \mid E \cap U = 0) \left( 1 - \gamma u \right), \,\,\, u \in \{ 0, 1\}. \label{eq.exposure.model.3}
\eeq
It is easily verified that 
$$
\pi_{E,U}(u) = P(y = 1 \mid E \cap U = u) \,\,\, u \in \{ 0, 1\}.
$$ 
This formulation as a compound logistic response function allows a separation between covariates predictive of exposure and
covariates predictive of vaccine efficacy. Introducing the expressions of Equation \eqref{eq.exposure.model.1} and \eqref{eq.exposure.model.2}
we have vaccine efficacy 
$$
\gamma = 1 - \frac{\tau_uP(E)}{\tau_{u^c} P(E)} =  1 - \frac{\tau_u}{\tau_{u^c}}.
$$  
Note that $\tau_u, \tau_{u^c}$ are probabilities conditioned on $E$. Then the model of Equation \eqref{eq.exposure.model.3} becomes, 
\beq
\pi_{E,U}(u) = \tau_{u^c} P(E) \left( 1 - \gamma u \right). \label{eq.exposure.model.4}
\eeq
We can replace  $u$ in \eqref{eq.exposure.model.4} with logistic response function $\phi(\eta_1)$. For clarity, we may set $\eta_1 = \alpha(z - z_{thr})$, possibly introducing additional covariates additively. We can introduce a second logistic response function, setting  $\gamma = \phi(\eta_2)$ in \eqref{eq.exposure.model.4}. Then any covariates in $\eta_2$ can be confined to those predictive specifically of vaccine efficacy. Note that 
$$
\phi(-\eta_2) = 1 - \phi(\eta_2) = \frac{\tau_u}{\tau_{u^c}}
$$
is then interpretable as a relative risk, expected to be strictly less than one.  We can then introduce a third logistic response function, setting  
$P(y=1 \mid U = 0)  = \tau_{u^c} P(E) = \phi(\eta_3)$.  Then $\eta_3$ can include covariates related to exposure, as well as covariates 
predictive of an unprotected individual's risk of infection given exposure.

\section{Inference for General Binary Models}\label{sec.bin.models}

Suppose we are given binary observations $y_i$ with mean $\pi_i$, $i = 1,\ldots, n$. Assume $\pi_i = \pi_i(\theta)$ for parameter $\theta \in \Theta \subset \real^p$.  The variance of $y_i$ is then $\tau_i = \tau_i(\theta) = \pi_i (1 - \pi_i)$, $i = 1,\ldots,n$. We use the following notation to denote the collection of partial derivatives (assumed to exist):
$$
\pip_{ij} = \pip_{ij}(\theta) = \pd{\pi_i}{\theta_j},  \,\,\, i = 1,\ldots,n, \,\,\, j = 1,\ldots,p.
$$
Let $\vecy$, $\vecpi$, $\vectau$, $\vecpip_j$, be $n$-dimensional column vectors with $i$th element $y_i$, $\pi_i$, $\tau_i$, $\pip_{ij}$, respectively.  For any $n$-dimensional (column) vector $\vecv$ let $\diag{\vecv}$  be the $n \times n$ diagonal matrix with diagonal $\vecv$. Then $I_r$ is the $r \times r$ identity matrix. For a square matrix $A$, let $A^{1/2}$ be any matrix satisfying $A^{1/2} A^{1/2} = A$.

Since the marginal variances of $\vecy$ are given by $\vectau$, the covariance matrix of $\vecy$ can be written 
\beq
V_{\theta} = \diag{\vectau(\theta)^{1/2}}   R_{\vecy}  \, \diag{\vectau(\theta)^{1/2}}, \label{eq.cov.y}
\eeq
where $R_{\vecy}$ is the correlation matrix of $\vecy$. That correlation is unrelated to independent variables used in a predictive model, or to unknown parameters more generally, is considered reasonable, so we assume that  $R_{\vecy}$ does not depend on $\theta$. In this case $V_{\theta}$ depends on $\theta$ only through $\vectau(\theta)$. 

\subsection{Estimating equations}

Under the assumption of independent responses the log-likelihood function for $\theta$ is  
\beq
L(\theta)  =  \sum_{i=1}^n y_i \log( \pi_i) + (1-y_i)   \log(1 -  \pi_i), \label{ll.function}
\eeq
and standard methods, such as Fisher scoring,  may be used to derive maximum likelihood estimate
$
\hattheta_{MLE} = \argmax_\theta \, L(\theta),
$
and standard error estimates.  

However,  if we have subject-level repeated measures  the derivation of the true likelihood, which correctly models response correlation,  is usually not straightforward.   In this case the estimating equation approach may be used, which depends on the distribution of $\vecy$ only through the first and second order moments. 

Suppose we are given an $n$-dimensional column vector 
$$
\tg(\theta) = [ g_1(\theta) \hdots g_n(\theta) ].
$$
Define the $n \times p$ matrix differential operator
$$
\Gamma(\tg) = \begin{bmatrix}  \pd{g_1(\theta)}{\theta_1} & \hdots & \pd{g_1(\theta)}{\theta_j}  & \hdots & \pd{g_1(\theta)}{\theta_p} \\ 
\vdots && \vdots && \vdots \\
\pd{g_n(\theta)}{\theta_1} & \hdots & \pd{g_n(\theta)}{\theta_j}  & \hdots & \pd{g_n(\theta)}{\theta_p} \end{bmatrix} 
$$
Estimating equations generally take the form 
\beq
G(\theta) =  \Gamma( \vecpi)^T  W_\theta^{-1} [ \vecy  - \vecpi ] = 0,  \label{qlik.eq2}
\eeq
where $W_\theta$ is a positive definite matrix for all $\theta$. It will be convenient to define the matrices:
\beas
\Omega^W_\theta &=& \Gamma(\vecpi)^T  W_\theta^{-1} \Gamma(\vecpi) \\
\Omega^V_\theta &=& \Gamma(\vecpi)^T  V_{\theta}^{-1} \Gamma(\vecpi) \\
\Omega^{WV}_\theta &=& \Gamma(\vecpi)^T  W_\theta^{-1} V_{\theta} W_\theta^{-1} \Gamma(\vecpi).
\eeas
Define solution $G(\hattheta) = 0$. This form of estimation is shared by many widely used methods. These include generalized estimating equations (GEE), which extend generalized linear models to allow correlated responses \citep{liang1986longitudinal}, or quasi-likelihood estimation, in which the relationship between mean and variance need not be constrained by any distributional assumptions \citep{wedderburn1974}. 

Conditions for asymptotic normality of $\hattheta$ for these various models are given in, for example, \cite{mccullagh1983} or \cite{wedderburn1974}. It is generally assumed that, with probability approaching 1 as $n \rightarrow \infty$, a unique solution to the system of equations \eqref{qlik.eq2} exists, or at least that a single solution in a neighborhood of the true parameter $\theta^*$ can be identified. 
More generally,  mathematical validation of these conclusions for estimating equations \eqref{qlik.eq2} for quite general models follows from Theorem 13.1 of \cite{almudevar2021theory}.  In this case, under conditions permitting application of Taylor's theorem, we have approximation
\beq
\hattheta - \theta^* \approx \dash [G^\pr(\theta^*)]^{-1} G(\theta^*), \label{eq.taylor}
\eeq
where $G^\pr(\theta)$ is the Jacobian matrix of $G(\theta)$ with respect to $\theta$. Further approximations rely on the observation that the elements of $G(\theta)$ and $G^\pr(\theta)$ are random sums. Therefore, under  general conditions we may use approximations $G^\pr(\theta^*) \approx E_{\theta^*}[G^\pr(\theta^*)]$ and  $G(\theta^*) \sim N(0, \Omega^{WV}_{\theta^*})$, following some matrix algebra. This gives the approximation
\beq
\hattheta - \theta^* \approx \dash E_{\theta^*}[G^\pr(\theta^*)]^{-1} G(\theta^*). \label{eq.taylor.2}
\eeq
In addition, we have  
\bea
E_{\theta^*}[G(\theta^*)] &=& 0, \nonumber \\
E_{\theta^*}[G^\pr(\theta^*)] &=& \dash\Omega^W_{\theta^*}. \label{eq.g.moments}
\eea 
This leads to approximate distribution
\beq
\hat{\theta} \sim N(\theta^*, \Sigma^W_{\theta^*}) \label{asymptotic.distribution.2}
\eeq
where 
\beq
\Sigma^W_\theta =  \left[ \Omega^W_{\theta}  \right]^{-1}  \Omega^{WV}_{\theta}  \left[ \Omega^W_{\theta}  \right]^{-1}. \label{eq.sig.w}
\eeq
If $W_\theta = V_{\theta}$, then  it is easily verified that
\beq 
\Sigma_\theta^W =  \left[ \Omega^V_{\theta}  \right]^{-1}. \label{def.sigma.theta}
\eeq
Then substitute $\hattheta$ for $\theta^*$, giving $\Sigma^W_{\theta^*} \approx \Sigma^W_{\hattheta}$. 

Finally, under given regularity conditions the solution  $\hat{\theta}$ can be obtained as the limit  of the  iterative 
algorithm
\beq
\theta_{n+1} = \theta_n + \left[ \Omega^W_{\theta_n} \right]^{-1} G(\theta_n),\,\,\ n \geq 0,  \label{eq.iterates.w}
\eeq
given some suitable initial solution $\theta_0$.

We may also wish to estimate fitted curves $\hatpi(\vecx)$ for arbitrary predictor values $\vecx$. Following \cite{cox1995} the relevant asymptotic variance is 
\beq
 \var{\hatpi(x)}  = \left[\pd{\pi(x)}{\theta}\right]^T    \Sigma^W_{\hattheta} \left[\pd{\pi(x)}{\theta}\right].  \label{eq.confidence.band}
\eeq
An approximation follows by substituting $\hattheta$ for $\theta$. If an $\alpha/2$ normal critical value is used to construct the confidence bands, then the confidence level $1-\alpha$ must be interpreted marginally. In \cite{cox1995} it is shown that a simultaneous $1-\alpha$ confidence band results when the critical value $(\chi^2_{p,\alpha})^{1/2}$ from a $\chi^2$ distribution with $p$ degrees of freedom is used.

\subsection{Estimating procedures}

So far, the only assumption imposed for weight matrix $W_\theta$ used in the estimating equations \eqref{qlik.eq2}  is positive definiteness and even this can be relaxed. However, the choice $W_\theta = V_{\theta}$ is provably optimal, and is therefore the obvious choice, absent other considerations. 
  
\subsubsection{Independent responses}\label{sec.alg.ind.mle}

If responses are independent, parameters are fit by maximizing the likelihood function  \eqref{ll.function}. 
This is equivalent to setting $W_\theta = V_{\theta} = D_{\vectau}$. Inference can be based on approximation with variance matrix \eqref{def.sigma.theta}, so that
$$
\hattheta \sim N\left(\theta^*, [\Omega^V_{\hattheta}]^{-1}\right)
$$
is the relevant approximate distribution. 

\subsubsection{Dependent responses}\label{sec.alg.nind.mle}

If responses are dependent a number of options are available. The maximum likelihood estimates, calculated under the assumption of independence, are still viable estimates, again setting  $W_\theta  = D_{\vectau}$, and an approximate density can be based on \eqref{asymptotic.distribution.2}. The difference is that $W_\theta \neq V_{\theta}$, and so the correct asymptotic distribution is approximated by 
\beq
\hattheta \sim N\left(\theta^*, \Sigma^W_{\hattheta}\right), \label{eq.normal.sigw}
\eeq
which requires an estimation  of  $V_{\theta}$, and therefore $R_{\vecy}$. This can be obtained by first solving the estimating equations  with $W_\theta  = D_{\vectau}$, yielding estimate $\hattheta_I$. Estimates of the mean and variance of each response $y_i$ are available as the elements of $\vecpi = \vecpi(\hattheta_I)$,     $\vectau = \vectau(\hattheta_I)$. The  Pearson residuals are then
\beq
\hatr_i = (y_i - \hatpi_i)/\hattau_i^{1/2}, \label{pearsons.correlation}
\eeq
from which correlation can be directly estimated, using, for example, the approach proposed in \cite{liang1986longitudinal} in the context of GEE estimation. The residuals contain information about $R_{\vecy}$, although an unstructured estimation of the entire correlation matrix is generally not feasible.  The standard approach is to assume a parametric form for $R_{\vecy}$, estimating the parameters from residuals $\hatr_i$.  

Once $R_{\vecy}$ is estimated, there are two options.  The original estimate $\hattheta_I$ can be accepted, with inference based on  \eqref{eq.normal.sigw}, employing $R_{\vecy}$ and Equation \eqref{eq.cov.y} in the calculation of $\Sigma^W_{\hattheta}$. However, since an estimate of $V_{\theta}$ is now available  we can solve estimating equations  with $W_\theta  = V_{\theta}$, yielding solution $\hattheta_R$ with an 
 asymptotic distribution approximated by 
\beq
\hattheta_R \sim N\left(\theta^*, [\Omega^V_{\hattheta}]^{-1}\right). \label{eq.normal.sigv}
\eeq

\subsection{Exchangeable correlation}\label{sec.exchangeable.correlation}

Under exchangeable correlation, random variates are partitioned into groups with common pairwise correlation $\rho$. Intergroup correlation is zero. Groups are typically defined as repeated measures from a common subject within a set of longitudinal data. This defines a parametric form for correlation matrix $R_{\vecy}$ which depends on a single parameter $\rho$ (Section \ref{sec.alg.nind.mle}). 

Methods for estimating $\rho$ are described in the literature.  Suppose sample indices $i = 1,\ldots,n$ are partitioned into groups $g = 1,\ldots,n_g$. Let $g_i$ identify the group of sample $i$. Then for any $i \neq i^\pr$ we have $\corr{y_i}{y_{i^\pr}} =  \rho \neq 0$ if $g_i = g_{i^\pr}$, and $\corr{y_i}{y_{i^\pr}} =  0$ otherwise. It is usually expected that $\rho > 0$, but this is not required. Following \cite{liang1986longitudinal}  we have an estimate of $\rho$ defined by
\beq
\hatrho = (N_{pair}-d)^{-1} \sum_{g=1}^{n_g} \sum_{\left\{i < {i^\pr} : g_i=g_{i^\pr}=g\right\}} \hatr_i \hatr_{i^\pr}, \label{eq.exchangeable.correlation}
\eeq 
where $N_{pair}$ is the total number of correlated pairs contributing to the sum, and $d$ is the number of estimated parameters.

\subsection{Regularization}

We will show below that there will be some advantage to stabilizing the model fit by using regularization. The ridge regression solution is to  minimize 
$$
\Lambda = SSE + \rrp D,
$$
where $SSE$ is the model error sum of squares, $\rrp$ is  a nonnegative penalty constant, $D$ is a penalty term given by 
$$
D = \theta^T B \theta, 
$$
$\theta$ is the $p \times 1$ parameter column vector, and $B$ is a $p \times p$ positive-definite matrix, which does not depend on $\theta$ or $\rrp$ \citep{hoerl1970ridge}. This procedure renders the fit more stable while shrinking the parameter estimates towards zero. For logistic regression with independent responses, $SSE$ is replaced with deviance. For dependent responses, the log-likelihood, and therefore the deviance is not formally defined. In this case, this form of regularization can be implemented directly into the estimating equations \eqref{qlik.eq2}, giving
\beq
G_{\rrp}(\theta) =  G(\theta) - \rrp \pd{D}{\theta} = 0.  \label{qlik.eq3}
\eeq
For independent responses, the solution to these equations is equivalent to 
$$
\hattheta_{MLE} = \argmin_\theta \,\mi 2L(\theta) + \rrp D, 
$$
but represents the equivalent penalized fitting criterion when the log-likelihood is not defined. Often, the choice $B = I_p$ is made, so that 
\beq
D = \sum_{j=1}^p \theta_j^2,\mbox{ and } \pd{D}{\theta} = 2 \theta. \label{eq.penalty}
\eeq
The asymptotic distribution of $\hattheta$ can be deduced using the same method as that based on Equation  \eqref{eq.taylor.2}, with the first and second order moments  of Equation \eqref{eq.g.moments} replaced by 
\bea
E_{\theta^*}[G_{\rrp}(\theta^*)] &=& \dash  {\rrp} \theta^*, \nonumber \\
E_{\theta^*}[G_{\rrp}^\pr(\theta^*)] &=& \dash \Omega^W_{\theta^*} - {\rrp} I_p.\label{eq.glambda.moments}
\eea 
This gives 
\bea
\hattheta &\approx& \theta^*  + \left[ \Omega^W_{\theta^*} + {\rrp} I_p \right]^{-1} 
\left[ G(\theta^*) - {\rrp} \theta^* \right] \nonumber  \\
&=&  \left[ \Omega^W_{\theta^*} + {\rrp} I_p \right]^{-1}   \left[ \Omega^W_{\theta^*} \theta^* + 
G(\theta^*) \right], \label{eq.iterates.w.lambda}
\eea
leading to approximation $\hattheta \sim N(\theta^{\rrp}_0, \Sigma^{W,{\rrp}}_{\theta^*})$ based on the approximate moments
\bea
\theta^{\rrp}_0 &=& E_{\theta^*}[\hattheta] \approx  \left[ \Omega^W_{\theta^*} + {\rrp} I_p \right]^{-1}    \Omega^W_{\theta_0} \theta_0  \nonumber \\
\Sigma^{W,{\rrp}}_{\theta_0} & = &  \vars{\hattheta}{\theta_0} \approx 
\left[ \Omega^W_{\theta_0} + {\rrp} I_p \right]^{-1}  \Omega^{WV}_{\theta_0}
\left[ \Omega^W_{\theta_0} + {\rrp} I_p \right]^{-1}, \label{eq.ridge.moments}
\eea
(essentially this argument is given in \cite{le1992ridge}). Inclusion of $D$ induces bias, and usually smaller variance, therefore the choice of ${\rrp}$ represents a bias-variance trade-off. Selection of ${\rrp}$ by cross-validation is discussed in the next section. 

\subsection{Model selection by cross-validation}\label{sec.cross.validation}

Cross validation may be used for variable or model selection, and to select  the penalty constant ${\rrp}$
defined in \eqref{eq.penalty}.  The strategy is to  randomly  partition the complete data set (of $n$ observations) into $K$ groups (as evenly as possible). Each group separately is taken as test data,  then $\hattheta_{train}$ is estimated  using the training data (the complete data set after deleting the test data). Then for each response $y_{test}$ in the test data,  the fitted value $\haty_{test}$ is calculated based on  estimate $\hattheta_{train}$ and paired predictor $\tx_{test}$.  

After  applying the procedure to the $K$ test groups, each response $y_i$ is paired with a cross-validated fitted value $\haty_i^*$.  The final step is to calculate a goodness-of-fit score 
$$
 GOF = GOF(y_1,\ldots,y_n; \haty_1^*, \ldots, \haty_n^*), 
$$
 which measures collectively the distances between $y_i, \haty^*_i$.  Possible goodness-of-fit scores include deviance:
$$
dev = \sum_{i=1}^n y_i \log(\haty^*_i) +   (1-y_i) \log(1-\haty^*_i),
$$
area under ROC curve: 
$$
AUC = \frac{\sum_{i \in I_0} \sum_{i^\pr \in I_1} I\{  \haty^*_{i^\pr} > \haty^*_{i} \} + 0.5 \times  I\{  \haty^*_{i^\pr} = \haty^*_{i} \}}{ |I_0| \times |I_1|},  
$$
where $I_j$ are the indices for which $y_i = j$, $j = 0,1$, and the error sum of squares defined by the quadratic: 
$$
SSE = \bar{y}^T V_{\theta}^{-1} \bar{y}
$$
where  $\bar{y}$ is an $n\times 1$ column vector with $i$th row equal to $y_i - y^*_i$, $i = 1,\ldots,n$, and $V_{\theta}$ is estimated using the complete data set. 

 In our implementation, ${\rrp}$  can be varied over a grid spanning ${\rrp} = 0$ to some upper bound $\rrp^{max}$. For each proposed ${\rrp}$ a model is fit, and evaluated using cross-validation. The choice $\rrp^{GOF}$ is the one optimizing whatever $GOF$ score is used. 
 
\section{Compound Logistic Model}\label{sec.compound.logistic.models}

We next show how the elements $\vecpi(\theta)$, $\vectau(\theta)$, $\vecpip_j(\theta)$ and the derived components $\Gamma(\vecpi(\theta))$, $V_{\theta}$, $\theta \in \real^p$,  are specialized to the compound logistic model.   Suppose each mean binary response element $\pi_i(\theta)$ of  $\vecpi(\theta)$ is constructed from $m$ components $\pi_{ik}(\theta)$, $k = 1,\ldots,m$ via compounding function $H:\real^m \rightarrow (0,1)$, giving
$$
\pi_i = H(\pi_{i1}, \ldots, \pi_{im}), \,\,\, i = 1,\ldots, n.
$$
Write the following partial derivatives as:
$$
H^\pr_k(u_1,  \ldots, u_m)   =  \pd{H(u_1, \ldots, u_m)}{u_k},\,\,\, k = 1,\ldots,m. 
$$
Then for convenience denote 
$$
H^\pr_{ik} =  H^\pr_{ik}(\theta) = H^\pr_k(\pi_{i1}(\theta), \ldots, \pi_{im}(\theta)),
$$
and construct $n$-dimensional column vectors $\vecHp_k = \vecHp_k(\theta) = [H^\pr_{1k} \hdots H^\pr_{nk}]$, 
and $\vecpi_k = \vecpi_k(\theta) =  [\pi_{1k} \hdots \pi_{nk}]$,    $k = 1,\ldots,m$. The derivative of response $\pi_i(\theta)$ with respect to $\theta_j$ is then decomposed into 
\beq
\pd{\pi_i(\theta)}{\theta_j} = \sum_{k=1}^m H^\pr_{ik} \pd{\pi_{ik}(\theta)}{\theta_j},   \label{eq.compound.model.1}
\eeq
from which it follows that 
\beq
\Gamma(\vecpi) = \sum_{k=1}^m D_{\vecHp_k} \times  \Gamma(\vecpi_k).  \label{eq.compound.model.2}
\eeq
Interestingly, the gradient matrix of the estimating equation \eqref{qlik.eq2} for the model $\vecpi(\theta)$ is simply a weighted sum of the gradient matrices for the models   $\vecpi_k(\theta)$ considered separately. 

\subsection{Application to Logistic Response Functions}

We can develop \eqref{eq.compound.model.2} further in the context of logistic regression.  Suppose with each of the components we associate 
an $n \times p$ design matrix $X_k$. Then $\pi_{ik} = \phi(\eta_{ik})$, $\eta_{ik} = \theta^T \tx_{ik}$, where $\tx_{ik}$ is the $i$th row of $X_k$. 
Then
$$
 \pd{\pi_{ik}(\theta)}{\theta_j} = \pd{\eta_{ik} }{\theta_j}   \phi(\eta_{ik}) (1 - \phi(\eta_{ik})) = x_{ijk} \tau_{ik}
$$
where $\{ X_k \}_{ij} = x_{ijk}$ and $\tau_{ik} = \pi_{ik}(1 - \pi_{ik})$. We may then write 
\beq
\Gamma(\vecpi_k)  = D_{\vectau_k} X_k, \,\,\, k = 1,\ldots,m, \label{eq.gamma.compound}
\eeq
so that \eqref{eq.compound.model.2} simplifies to 
\beq
\Gamma(\vecpi) = \sum_{k=1}^m D_{\vecHp_k}  D_{\vectau_k} X_k.  \label{eq.compound.model.3}
\eeq

Note that  we have assumed that each component $\pi_{ik}(\theta)$, $k = 1,\ldots, m$,  is a function of a common parameter $\theta$. In an application it may be more natural to assign distinct parameters to each component. In this case, it would simply be a matter of partitioning the elements of $\theta$, so that each element $\theta_j$ appears in only one component. Otherwise, an element may appear in more than one component. The methodology remains the same in either case. This issue is discussed further below.

\subsection{Example (Asymptote Model)}\label{sec.ex.asymptote.model}

We consider the asymptote model  of \eqref{eq.general.cop.etac} - \eqref{eq.general.cop.etac.H}, with independent responses. 
Regarding the estimating equation  \eqref{qlik.eq2} and algorithm \eqref{eq.iterates.w} we will set $W_\theta = V_{\theta} = D_{\vectau(\theta)}$.

Suppose the component $\phi_1(\eta_1)$ incorporates a biomarker $z_i$, and $\phi_2(\eta_2)$,  $\phi_3(\eta_3)$  each incorporate a common covariate $x_i$. In addition, each component incorporates an independent intercept term, so we may write
\bea
\eta_{i1} &=& \beta_{0,1}  +    \beta_{1,1} z_i, \nonumber \\
\eta_{i2} &=& \beta_{0,2}  +    \beta_{1,2} x_i, \nonumber \\
\eta_{i3} &=& \beta_{0,3}  +    \beta_{1,3} x_i. \label{eq.ex.asymptote.model.eta}
\eea
The parameter is then $\theta = (\beta_{0,1},  \beta_{1,1}, \beta_{0,2},  \beta_{1,2}, \beta_{0,3},  \beta_{1,3}) \in \real^p$, $p = 6$.
The design matrices are then
$$
X_1 = \begin{bmatrix}  1 & z_1 & 0 & 0 & 0 & 0 \\ \vdots & \vdots & \vdots & \vdots & \vdots & \vdots \\
 1 & z_n & 0 & 0 & 0 & 0 \end{bmatrix},\,\,\,  X_2 = \begin{bmatrix}  0 & 0 & 1 & x_1 & 0 & 0 \\ \vdots & \vdots & \vdots & \vdots & \vdots & \vdots \\  0 & 0 & 1 & x_n & 0 & 0 \end{bmatrix}, 
$$ 
$$
X_3 = \begin{bmatrix}  0 & 0 & 0 & 0 & 1 & x_1 \\ \vdots & \vdots & \vdots & \vdots & \vdots & \vdots \\  0 & 0 & 0 & 0 & 1 & x_n  \end{bmatrix}.
$$ 
The six parameters of $\theta$ are assumed linearly independent. However, by \eqref{eq.compound.model.3} the gradient matrix 
$\Gamma(\vecpi)$ is a weighted sum of  $X_0$, $X_1$ and $X_2$, which have common column vectors. It is therefore important 
to verify that $\Gamma(\vecpi)$ is a full rank matrix (so that $\Omega^W_\theta = \Omega^V_\theta$ in algorithm  \eqref{eq.iterates.w} is invertible).  Directly from \eqref{eq.compound.model.3} we have $n \times p$ matrix
$$
\Gamma(\vecpi) = \begin{bmatrix}  
H^\pr_{11} \tau_{11} &   H^\pr_{11} \tau_{11} z_1 & 
H^\pr_{12} \tau_{12} &   H^\pr_{12} \tau_{12} x_1 & 
H^\pr_{13} \tau_{13} &   H^\pr_{13} \tau_{13} x_1 \\ 
\vdots & \vdots & \vdots & \vdots & \vdots & \vdots \\
H^\pr_{i1} \tau_{i1} &   H^\pr_{i1} \tau_{i1} z_i & 
H^\pr_{i2} \tau_{i2} &   H^\pr_{i2} \tau_{i2} x_i & 
H^\pr_{i3} \tau_{i3} &   H^\pr_{i3} \tau_{i3} x_i \\ 
\vdots & \vdots & \vdots & \vdots & \vdots & \vdots \\
H^\pr_{n1} \tau_{n1} &   H^\pr_{n1} \tau_{n1} z_n & 
H^\pr_{n2} \tau_{n2} &   H^\pr_{n2} \tau_{n2} x_n & 
H^\pr_{n3} \tau_{n3} &   H^\pr_{n3} \tau_{n3} x_n \end{bmatrix}. 
$$
From Equation \eqref{eq.general.cop.etac.H} we have partial derivatives
$$ 
\pd{H(u_1, u_2, u_3)}{u_1} =  u_3 - u_2;\,\pd{H(u_1, u_2, u_3)}{u_2} =  1 - u_1;\,\pd{H(u_1, u_2, u_3)}{u_3} =   u_1.
$$
This gives
$$
H^\pr_{i1} = \pi_{i3} - \pi_{i2};\, H^\pr_{i2} = 1 - \pi_{i1};\, H^\pr_{i3} = \pi_{i1}.
$$
To simplify the demonstration, we will assume $x_i$ is an indicator variable, with $x_i = 1$, $i = 1, \ldots, m$ and  $x_i = 0$, $i = m+1, \ldots, n$. We can identify values  $\pi_{i2}= \pi_{2}^+, \pi_{2}^-$ and $\pi_{i3}= \pi_{3}^+, \pi_{3}^-$ that hold for $x_i = 1, 0$, respectively. Set $d^+ =  \pi_{3}^+ - \pi_{2}^+$ and $d^- =  \pi_{3}^- - \pi_{2}^-$.  Also set $\tau_{k}^+ =  \pi_{k}^+ (1 -  \pi_{1}^+)$, $\tau_{k}^- =  \pi_{k}^- (1 -  \pi_{k}^-)$, $k = 1,2,3$.   The quantities $\pi_{k}^+$, $\pi_{k}^-$ are in the interval $(0,1)$, so that  $\pi_{k}^+$, $\pi_{k}^-$, $\tau_{k}^+$, $\tau_{k}^-$ are all strictly positive. Let $\tilde{0}$, $\tilde{1}$ be $n$-dimensional column vectors with elements zero and one, respectively.  Finally, a partition of any  $n$-dimensional column vector $\tilde{u}$ into elements $1, \ldots, m$ and elements $m+1, \ldots, n$ will be denoted  $(\tilde{u}^+, \tilde{u}^-)$. Extend this notational convention to any matrix with $n$ rows.  

We may then write
\begin{flushleft}
$\Gamma(\vecpi) = $
\end{flushleft}
$$
\left[ \begin{array}{cccccc} d^+ \vectau_1^+ &
d^+ (\vectau_1 \circ \tilde{z})^+ &
\tau_{2}^+ (\tilde{1} \ng  \vecpi_1^+)  &
\tau_{2}^+ (\tilde{1} \ng \vecpi_1^+)  &
\tau_{3}^+ \vecpi_1^+  &
\tau_{3}^+ \vecpi_1^+  \\
\rule[2pt]{25pt}{0.5pt} & \rule[2pt]{45pt}{0.5pt} & \rule[2pt]{45pt}{0.5pt} & \rule[2pt]{45pt}{0.5pt} &
\rule[2pt]{25pt}{0.5pt} & \rule[2pt]{25pt}{0.5pt} \\
d^- \vectau_1^-  &
d^- (\vectau_1 \circ \tilde{z})^-   &
\tau_{2}^-  (\tilde{1} \ng \vecpi_1^-)  &
\tilde{0} &
\tau_{3}^-   \vecpi_1^-  &
 \tilde{0} \end{array}\right],
$$
where $\circ$ is the Hadamard product. Assume $m \geq 4$, $n-m \geq 4$. Then suppose the vectors $B^+ = (\tilde{1}^+, \tilde{z}^+, \vecpi_1^+)$ are linearly  independent vectors in $\real^m$. 
Then $\vectau_1^+ = \vecpi_1^+ \circ (\tilde{1}^+ -  \vecpi_1^+)$, and so is a second order interaction with respect to vectors in $B^+$. 
Similarly, $(\vectau_1 \circ \tilde{z})^+ = \vecpi_1^+ \circ (\tilde{1}^+ -  \vecpi_1^+) \circ \tilde{z}^+$, and so is a third order interaction 
with respect to vectors in $B^+$. It follows that $V^+ = (\tilde{1}^+, \vecpi_1^+, \vectau_1^+,  \vectau_1^+ \circ \tilde{z}^+)$ are linearly independent vectors in $\real^m$. 
We then suppose  $B^- = (\tilde{1}^-, \tilde{z}^-, \vecpi_1^-)$ are linearly  independent vectors in $\real^{n-m}$, and similarly conclude that 
 $V^- = (\tilde{1}^-, \vecpi_1^-, \vectau_1^-,  \vectau_1^- \circ \tilde{z}^-)$ are linearly independent vectors in $\real^{n-m}$. 

Next, let $C_i$, $i = 1, \ldots, 6$,  be the column vectors of $\Gamma(\tg)$, with partitions $(C_i^+, C_i^-)$. Suppose there are real numbers $a_i$ for which
\beq
\sum_{i=1}^6 a_i C_i = \tilde{0}. \label{eq.ac6}
\eeq
If either $d^+ \neq 0$ or $d^- \neq 0$ we must have $a_1 = a_2 = 0$ in  \eqref{eq.ac6}. Then since $C_4^- = C_6^- = \tilde{0}^-$, and $C_3^-$, $C_5^-$ are linearly independent, we must also have  $a_3 = a_5 = 0$. Finally, $C_4^+$, $C_6^+$ are linearly independent, so we must have $a_4 = a_6 = 0$. It follows that the column vectors of $C_1, \ldots, C_6$  are linearly independent so that $\Gamma(\vecpi)$ is full rank.

%%%\Omega^V_\theta &=& \Gamma(\vecpi)^T  V_{\vecy}^{-1} \Gamma(\vecpi) \\

\section{Fitting Methods}\label{sec.fitting.methods}

The methods introduced here are supported in the \ttt{R}-package \ttt{CLmodel}. Fitting options
are available as given in Table \ref{tab.fit.options} below.

\begin{table}\caption{\label{tab.fit.options} Fitting options available in \ttt{CLmodel} \ttt{R}-package.}
\centering
\fbox{%
\begin{tabular}{lcl} \hline
Option & Values & Description \\\hline
\ttt{cop.flag} & \ttt{TRUE/FALSE} & Force positive biomarker gradient \\
\ttt{exchangeable}  & \ttt{TRUE/FALSE} & Use exchangeable correlation model  \\
\ttt{quasi.lik} & \ttt{TRUE/FALSE}  & Use quasi-likelihood estimation \\
\ttt{lambda} & $\lambda \geq 0$ & Regularization parameter \\
& &  ($\lambda = 0.0$ for no regularization) \\  \hline
\end{tabular}}
\end{table}

The \ttt{exchangeable} option is set to \ttt{FALSE} or \ttt{TRUE} according to whether the responses are independent or possess exchangeable correlation, as described
in Section \ref{sec.exchangeable.correlation}.  In the latter case, the groups must be supplied.  

For the independent response case, the true variance is $V_{\theta} = D_{\vectau(\theta^*)}$ so set $W_\theta =  D_{\vectau(\theta)}$ in estimating equation \eqref{qlik.eq2}, 
asymptotic distribution \eqref{asymptotic.distribution.2} and iterative algorithm \eqref{eq.iterates.w}. This is equivalent to the MLE. 
 
If the option \ttt{exchangeable = TRUE} is selected  then the MLE (under the independence assumption) will be calculated (that is, using $W_\theta =  D_{\vectau(\theta)}$ in 
iterative algorithm \eqref{eq.iterates.w}). The resulting fit is used to give an estimate of the correlation coefficient $\hat{\rho}$, as in Equation  \eqref{eq.exchangeable.correlation}. This estimate in turn is used to construct the estimated correlation matrix $\hat{R}_{\ty}$, and approximate variance matrix $V_\theta \approx \hat{V}_\theta = D_{\vectau(\theta)^{1/2}} \hat{R}_{\ty} D_{\vectau(\theta)^{1/2}}$.

At this point there are two further options. If options \ttt{exchangeable = TRUE} and  \ttt{quasi.lik = FALSE}  are selected, then the parameter estimates obtained by setting  $W_\theta =  D_{\vectau(\theta)}$ in iterative algorithm \eqref{eq.iterates.w} are accepted, but the standard errors are obtained by setting $V_\theta \approx \hat{V}_\theta$ in  asymptotic distribution \eqref{asymptotic.distribution.2}. If options \ttt{exchangeable = TRUE} and  \ttt{quasi.lik = TRUE}  are selected, then we set $W_\theta =   \hat{V}_\theta$  in  iterative algorithm \eqref{eq.iterates.w} and recalculate the parameter estimates. Then $W_\theta =  V_\theta =  \hat{V}_\theta$ is used in  asymptotic distribution \eqref{asymptotic.distribution.2} to derive standard errors. 

If the option \ttt{lambda = 0.0} no regularization is used. Otherwise iterative algorithm \eqref{eq.iterates.w} and asymptotic distribution  \eqref{asymptotic.distribution.2}  
are replaced with  \eqref{eq.iterates.w.lambda} and \eqref{eq.ridge.moments}, respectively, with $\lambda$ equal to the value specified in the option. 
 
\subsection{Transformation of biomarker gradient}\label{sec.biomarker.gradient} 
 
Recall the discussion regarding identifiability of Section \ref{sec.ex.asymptote.model}. It was shown that a necessary condition for identifiability is that the gradient associated with the designated biomarker is not zero. Suppose in the asymptote model of Section \ref{sec.ex.asymptote.model} we have $\eta_1 = \beta_{0,1} + \beta_{1,1} x_1$,
and that $x_1$ is the designated biomarker. In most applications the direction of the effect of $x_1$ on the response (equivalently, the sign of the $\beta_{1,1}$) would be known, or could be confidently inferred without resort to a formal estimate of  $\beta_{1,1}$. In this case, we can assume $\beta_{1,1} > 0$, reversing the sign of $x_1$ if needed. While it is possible to build such a constraint into  an optimizer, we opt instead to reparametrize $\eta_1 = \beta_{0,1} + \beta_{1,1} x_1$ to $\eta_1 = e^\xi (x_1 - \beta^*_{0,1})$, $\xi \in (-\infty, \infty)$. If additional covariates $x_2, \ldots, x_q$ are used in this component we can set $\eta_1 = e^\xi (x_1 - \eta_1^*)$, where $\eta_1^* = \beta^*_{0,1} + \beta^*_{2,1} x_2 + \ldots + \beta^*_{q,1} x_q$.  This would change the method of calculating $\Gamma(\vecpi)$ in estimating equation \eqref{qlik.eq2}, but the other components of the iterative algorithm and the asymptotic distribution would remain unchanged following that revision. The overall approach remains mathematically validated by Theorem 13.1 of \cite{almudevar2021theory}. 

\subsection{Initial solution}\label{sec.initial.solution} 

The discussion regarding identifiability in Section \ref{sec.ex.asymptote.model} makes clear that using a good initial solution for iterative algorithm \eqref{eq.iterates.w} or \eqref{eq.iterates.w.lambda} is important. In particular, we cannot start with a null model defined by a zero biomarker gradient.   The approach advocated here is to first fit the four parameter model based on \eqref{eq.general.cop.eta}, with single biomarker covariate $x$ and binary response $y$, setting $\eta = \alpha (x - x_{thr})$. This yields initial estimate $\htheta^*_0 = (\hat{\alpha}, \hat{x}_{thr}, \hat{a}_L, \hat{a}_U)$.  These parameters can then be transformed 
to the correct model parametrization. 

\section{Simulation Study}\label{sec.sim}

In this section we use a model simulation to assess the accuracy of the proposed methodology.  Consider the model of Section \ref{sec.ex.asymptote.model}, based on compounding function  \eqref{eq.general.cop.etac} - \eqref{eq.general.cop.etac.H}. We take the prediction terms of Equation \eqref{eq.ex.asymptote.model.eta}, where $z_i$ is the biomarker, and $x_i$ is some binary covariate. Following Section \ref{sec.biomarker.gradient},  we set $\eta_{i1} = \exp(\alpha)(z_i - \beta_{0,1})$, while   $\eta_{i2}$,  $\eta_{i3}$ remain as they are in   \eqref{eq.ex.asymptote.model.eta}. 

\begin{figure}
\centering
\includegraphics[height=2.0in, width=3.5in,viewport = 50 115 400 340, clip]{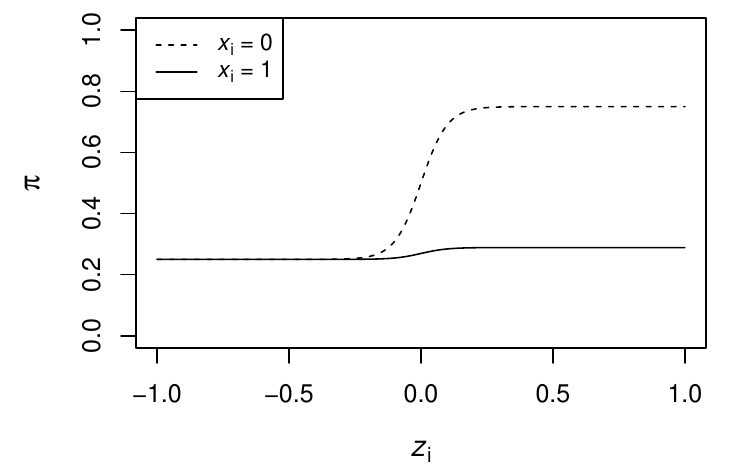}
\caption{Response curve for simulated model of Section \ref{sec.sim}.}\label{figTrueModel01}
\end{figure}

We set the total number of responses $n = 200$. In addition, we will impose an exchangeable correlation model with $\rho = 0.1$,   with observations randomly partitioned into groups of 5.  
The biomarker $z_i$ is uniformly simulated from interval $[-1,1]$. Each $x_i$ is randomly set to one with probability 0.5. Then set true parameter 
\beas
\theta_{true} &=& (\alpha, \beta_{0,1}, \beta_{0,2},\beta_{1,2},\beta_{0,3},\beta_{1,3}) \\
&=& (3, 0, \log(0.25/0.75), 0, \log(0.75/0.25), -2). 
\eeas
The response curve is shown in Figure \ref{figTrueModel01}. $N_{sim} = 50,000$ response vectors were simulated (the same covariates $z_i$, $x_i$ were used for each replication). Fit options were set to \ttt{cop.flag = TRUE}, 
\ttt{exchangeable = TRUE} and  \ttt{quasi.lik = TRUE} (Table \ref{tab.fit.options}). Using common replications, $N_{sim}$ fitted models were calculated with  $\lambda = 0.0$ (not regularized) and then
with  $\lambda = 0.1$ (regularized). 

%\subsection{Parameter estimation}
That ridge regression improves the stability of the numerical algorithm is clearly seen, in that 55.6\% of fits failed to converge for ${\rrp} = 0$, while all converged for  ${\rrp} = 0.1$.  Table \ref{tab.sim.study} gives numerical summaries for parameter estimates $\hattheta_k$ and reported standard errors $S_{\hattheta_k}$ for each parameter.  

We first note that the high convergence failure rate for the ${\rrp} = 0$ case appears to have a number of effects on the distribution of the estimates and their standard errors. The highest quantile of the standard errors, $Q_{0.99}$,  is uniformly larger for the ${\rrp} = 0$ case, especially for $\alpha$ and $\beta_{0,1}$, suggesting that regularization yields great stability and robustness. It should  also be noted that the standard deviations of the estimates for the ${\rrp} = 0$ case are uniformly smaller than for the ${\rrp} = 0.1$ case. This suggests that more divergent fits are being truncated by convergence failure.

\begin{figure}
\centering
\includegraphics[height=5.0in, width=5.0in,viewport = 10 10 500 500, clip]{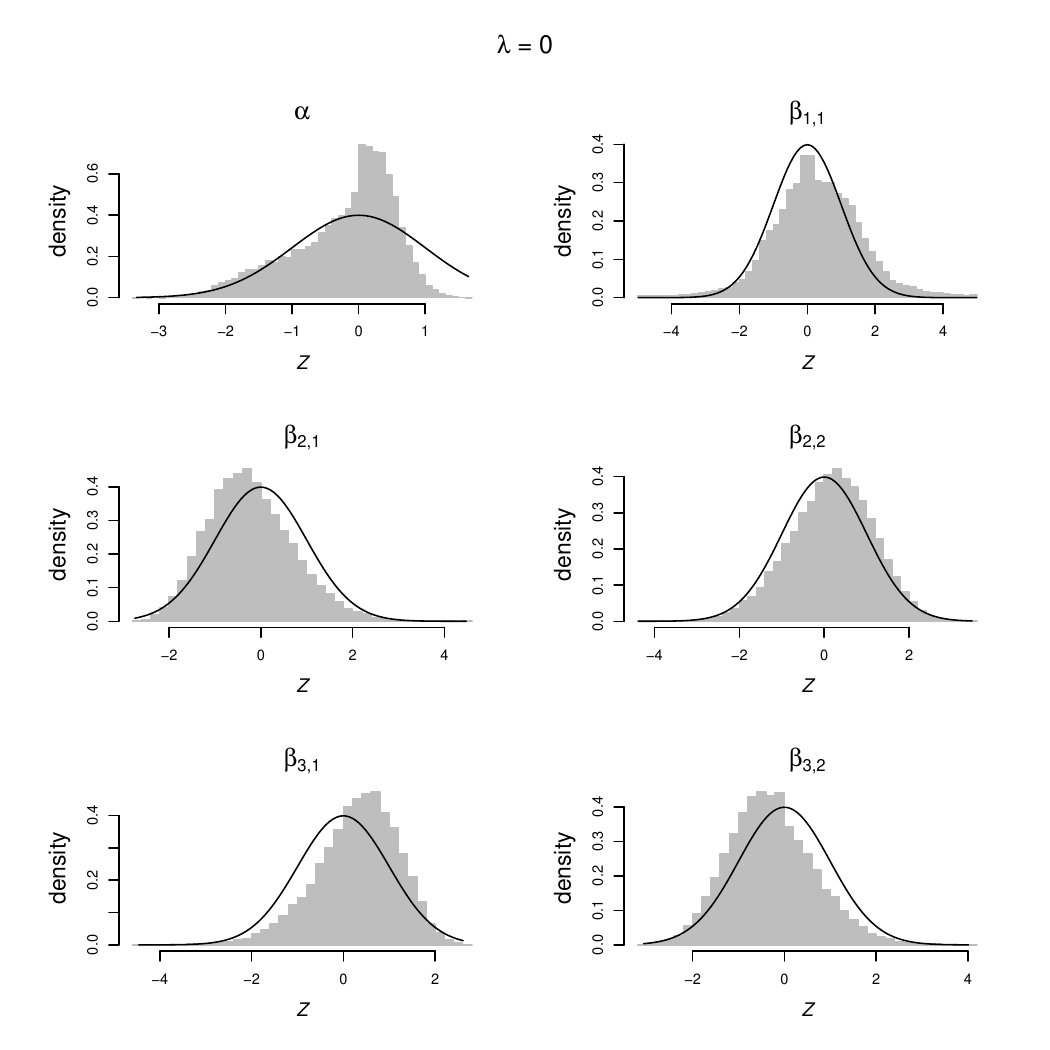}
\caption{Histograms of parameter estimate $Z$-scores $(\hattheta_j - \theta^*_j)/S_{\hattheta_j}$, from simulation study of Section \ref{sec.sim} (${\rrp} = 0$). Values $|Z| > 5$ are omitted.}\label{figSIM2a}
\end{figure}

\begin{figure}
\centering
\includegraphics[height=5.0in, width=5.0in,viewport = 10 10 500 500, clip]{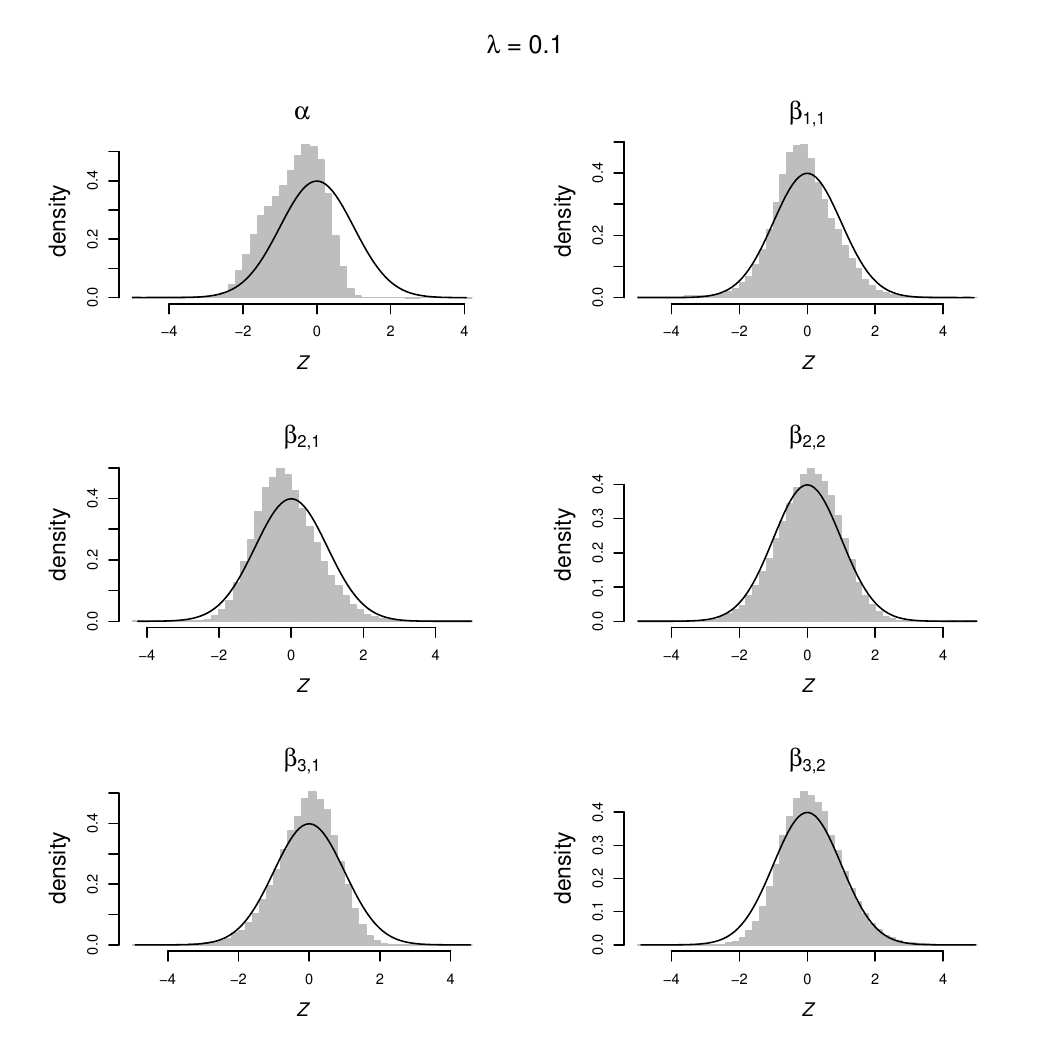}
\caption{Histograms of parameter estimate $Z$-scores $(\hattheta_j - \theta^*_j)/S_{\hattheta_j}$, from simulation study of Section \ref{sec.sim} (${\rrp} = 0.1$). Values $|Z| > 5$ are omitted.}\label{figSIM2b}
\end{figure}

Figures \ref{figSIM2a}-\ref{figSIM2b} give histograms of $Z$-scores 
$$
Z = \frac{\hattheta_j - \theta_j}{S_{\hattheta_j}}
$$
for parameters $j = 1,\ldots,5$. Values of $|Z| > 5$ were omitted from the histograms (frequencies are reported in Table \ref{tab.sim.study}).  Skewness is evident in most distributions. Nontheless, the actual coverage of a 95\% confidence interval based on the asymptotic normal distribution is close to 0.95 in all cases except for $\beta_{0,1}$ for the ${\rrp} = 0$ case.  

The estimates of $\rho = 0.1$ were reasonably accurate, with $\hat{\rho} = 0.087 \pm 0.063$ for ${\rrp} = 0$ and  $\hat{\rho} = 0.090 \pm 0.062$ for ${\rrp} = 0.1$. Ridge regression introduces bias, so that the estimated model will depend on the degree of penalization. However, from Table \ref{tab.sim.study} it can be seen that the parameter estimates for the two choices of ${\rrp}$ are comparable. 

% latex table generated in R 4.3.2 by xtable 1.8-4 package
% Tue Jan 27 16:05:30 2026
\begin{table}
   \caption{\label{tab.sim.study}Summary of simulation study of Section \ref{sec.sim}. Penalty constants ${\rrp} = 0, 0.1$ were both used. Sample means and standard deviations of each parameter estimate $\hattheta_j$ are  given. Various quantiles of the standard errors $S_{\hattheta_j}$  are also given. The entry ``Cover.'' gives the estimated actual coverage of a nominal 95\% confidence interval based on the asymptotic normal distribution. The entry $\% |Z| > 5$ gives the percentage of estimates which are further than 5 standard errors from the true parameter value.}
\centering
\fbox{%
\begin{tabular}{r|r|l|rrrrrr} \hline
&&& $\alpha$ & $\beta_{0,1}$ & $\beta_{0,2}$ & $\beta_{1,2}$ & $\beta_{0,3}$  & $\beta_{1,3}$ \\  \hline
${\rrp} = 0$ & $\hattheta_j$ & True value & 3.000 & 0.000 & -1.099 & 0.000 & 1.099 & -2.000 \\ 
&& Mean &  3.107 & 0.001 & -1.528 & 0.170 & 1.093 & -2.212 \\ 
 && SD &   1.088 & 0.144 & 4.345 & 1.173 & 4.644 & 1.336 \\ \cline{2-9}
&$S_{\hattheta_j}$ & $Q_{0.01}$&  0.581 & 0.013 & 0.295 & 0.414 & 0.293 & 0.377 \\ 
&& $Q_{0.25}$  &  0.944 & 0.049 & 0.356 & 0.489 & 0.372 & 0.458 \\ 
&& $Q_{0.5}$  &  1.239 & 0.074 & 0.393 & 0.529 & 0.419 & 0.499 \\ 
&& $Q_{0.75}$  &  1.706 & 0.109 & 0.458 & 0.594 & 0.505 & 0.574 \\ 
&& $Q_{0.99}$  & $> 10^6$  & $> 10^3$ & 3.806 & 3.957 & 4.834 & 4.805 \\ \cline{2-9}
&& Cover. &  0.973 & 0.857 & 0.968 & 0.957 & 0.972 & 0.961 \\ 
&& $\% |Z| > 5$ &  0.000 & 2.159 & 0.000 & 0.000 & 0.000 & 0.000 \\ \hline
${\rrp} = 0.1$ & $\hattheta_j$ & True value &   3.000 & 0.000 & -1.099 & 0.000 & 1.099 & -2.000 \\ 
&& Mean &    2.478 & -0.052 & -1.271 & 0.101 & 0.976 & -1.834 \\ 
 && SD &    5.439 & 3.445 & 8.060 & 8.408 & 8.902 & 7.916 \\ \cline{2-9}
&$S_{\hattheta_j}$ & $Q_{0.01}$&   0.582 & 0.029 & 0.305 & 0.418 & 0.299 & 0.371 \\ 
&& $Q_{0.25}$  &    0.860 & 0.077 & 0.362 & 0.490 & 0.367 & 0.444 \\ 
&& $Q_{0.5}$  &    1.009 & 0.107 & 0.401 & 0.530 & 0.407 & 0.479 \\ 
&& $Q_{0.75}$  &    1.174 & 0.150 & 0.478 & 0.599 & 0.476 & 0.538 \\ 
&& $Q_{0.99}$  &   1.813 & 0.432 & 1.075 & 1.048 & 0.929 & 0.896 \\ \cline{2-9}
&& Cover. &   0.950 & 0.951 & 0.974 & 0.965 & 0.971 & 0.961 \\ 
&& $\% |Z| > 5$ &   0.834 & 0.880 & 0.546 & 0.522 & 0.518 & 0.552 \\  \hline
\end{tabular}}
\end{table}

We finally note the difference in response between observations coded differently by indicator variable  $x_i$ evident in Figure \ref{figTrueModel01}. The difference is attributable to the parameter $\beta_{1,3} = -2$. The 
power for a two-sided alternative to null hypothesis $H_0 : \beta_{1,3} = 0$ at significance level $\alpha = 0.05$ was estimated to be 0.97, 0.98 for $\rrp = 0.0, 0.1$. 

\section{Example}\label{sec.examples}

The data used in this example were collected as part of the Midlife in the United States (MIDUS) project. This project originated in 1995 from a national survey by the MacArthur Foundation Research Network on Successful Midlife Development, which was extended into a longitudinal study in 2002 at the University of Wisconsin-Madison through a grant from the National Institute on Aging. The data include interview and questionnaire responses, biomarker measurements, and cognitive assessments of middle-aged adults in the United States, with the goal of identifying the relationships between physical, psychological, and social factors of well-being in adulthood. The collection of survey data for MIDUS 3, the most recent phase of the project and the basis of this paper's data, was conducted between 2013 and 2014, while the biomarker data were collected between 2017 and 2022. The results of the study have been made available through the Inter-University Consortium of Political and Social Research (ICPSR).

\begin{figure}
\centering
\includegraphics[height=3.5in, width=4.5in,viewport = 15 50 480 390, clip]{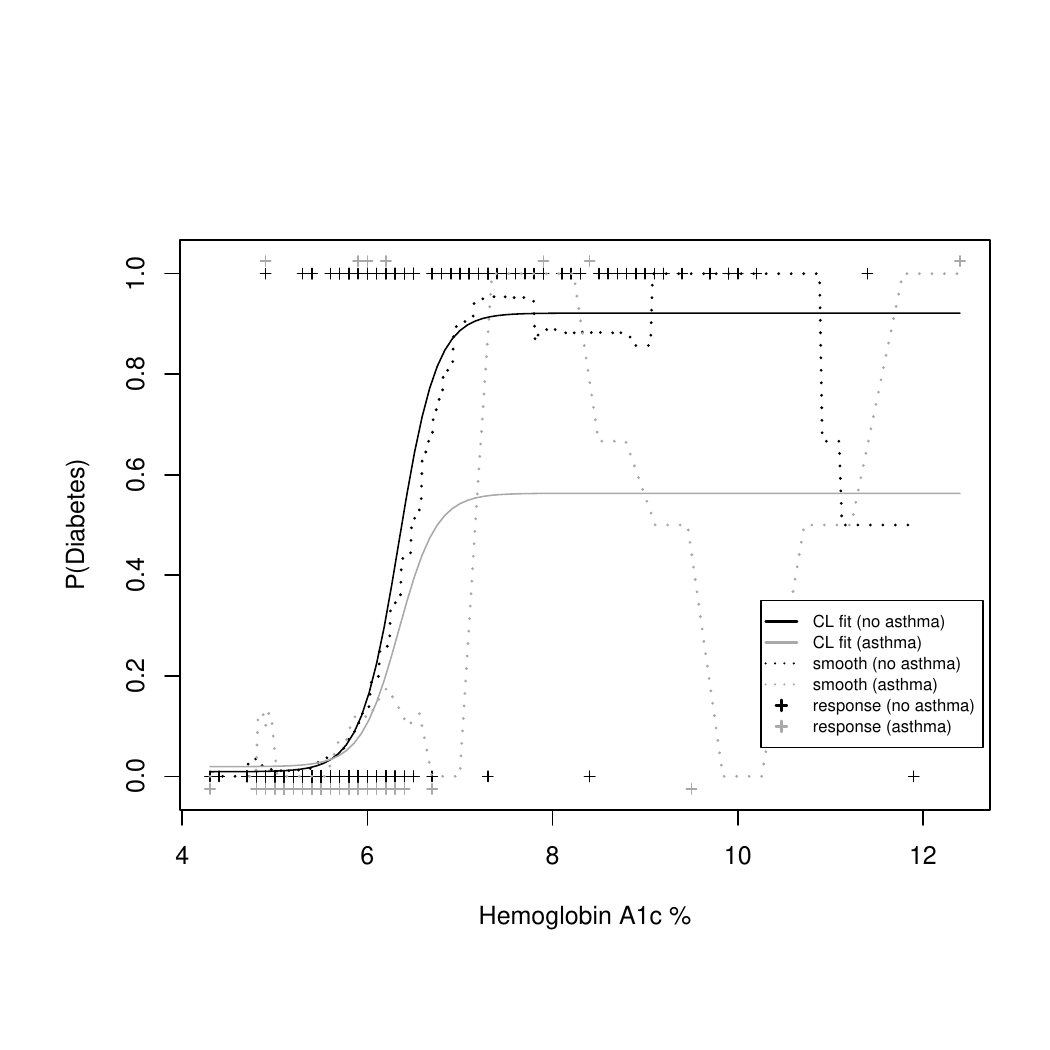}
\caption{Fitted response curves for selected COP model, Section \ref{sec.examples}}\label{figDiabetes1}
\end{figure}

\begin{figure}
\centering
\includegraphics[height=3.5in, width=4.5in,viewport = 15 50 480 390, clip]{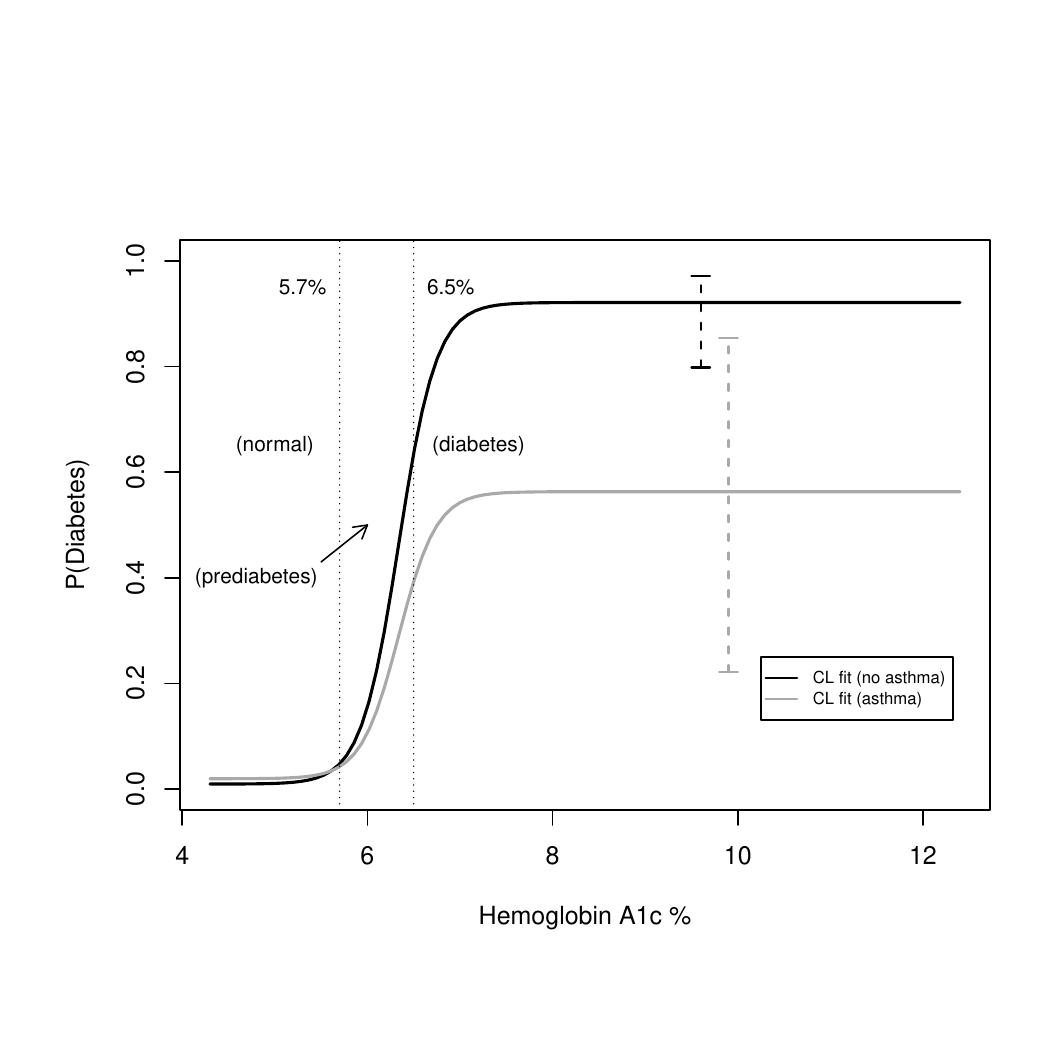}
\caption{Fitted response curves for selected COP model, Section \ref{sec.examples}}\label{figDiabetes2}
\end{figure}

Suppose we are given binary response $y_i$ which codes a subject with diabetes. Then $x_i$ is a binary covariate which codes a subject with asthma. We are then given a biomarker $z_i$ 
representing hemoglobin A1c\%, commonly used as a diagnostic biomarker for diabetes. We use the model of Section \ref{sec.ex.asymptote.model}, based on compounding function  \eqref{eq.general.cop.etac} - \eqref{eq.general.cop.etac.H}. Set $\eta_{i1} = \beta_{0,1} + \beta_{1,1} z_i$, $\eta_{i2} = \beta_{0,2} + \beta_{1,2} x_i$ and $\eta_{i3} = \beta_{0,3} + \beta_{1,3} x_i$. Using the \ttt{CLmodel} \ttt{R}-package 
with \ttt{exchangeable = FALSE} and \ttt{lambda = 0} (that is, the maximum likelihood estimate) we obtain the following output coefficient table:
\begin{verbatim}
             Est        SE          Z    p-val (Z)
b[1,1] -31.019268 5.9930032 -5.1759138 2.267982e-07
b[1,2]   4.893016 0.9797937  4.9939246 5.916451e-07
b[2,1]  -4.689845 0.9747431 -4.8113650 1.499030e-06
b[2,2]   0.758998 1.3542669  0.5604493 5.751730e-01
b[3,1]   2.463809 0.5539285  4.4478823 8.672101e-06
b[3,2]  -2.210098 0.9207106 -2.4004262 1.637599e-02
\end{verbatim}
Figures \ref{figDiabetes1} and \ref{figDiabetes2} present two views of the resulting fitted model. Figure \ref{figDiabetes1} shows the estimated response curve
$\hat{\pi}(z, x)$. Separate curves $\hat{\pi}(z, 0)$, $\hat{\pi}(z, 1)$ are plotted as a function of $z$ (hemoglobin A1c\%). In addition, kernel smoothers are superimposed, 
calculated separately for the $x = 0$, $x = 1$ cases. Finally, separate rug plots show the location and value of the individual responses. 

There are $n =  632$ observations, of which $n_1 = 89$ are coded as asthma positive. As expected, for small values of $z$ (an after-fasting measure of blood glucose levels) the 
response function (the probability of diabetes) is near zero,  while for larger values of $z$ the response function approaches separate asymptotes for the $x = 0$, $x = 1$ cases.
The difference in upper asymptotes is entirely attributable to the parameter $\beta_{1,3}$, here estimated to be   $\hbeta_{1,3} = -2.21$ (note the differing coefficient indexing conventions).  
From the table, $\beta_{1,3}$ differs significantly from zero with a $p$-value of 0.017. We conclude that the probability of diabetes at high levels of the biomarker is significantly lower for subjects 
with asthma (OR = 0.11; 95\% CI = (0.02, 0.69)).  

For the $x = 0$ (no asthma) group, the kernel smoother suggests that the  shape of the response curve is appropriate for the data, beginning with a near zero lower asymptote, followed by a rapid increase around $z = 6\%$, 
ending with an upper asymptote of approximately 0.92 (with a 95\% upper confidence bound of 0.97).  The rug plots indicates that data for the $x = 1$ (asthma) group is very sparse above $z = 6\%$. However, the kernel smoother 
similarly conforms to the estimated response curve for that group also. 

Figure \ref{figDiabetes2} shows the same fitted curves, and includes  $95\%$ confidence intervals (CI) for the upper asymptotes for the $x = 0$ and $x = 1$ groups. The two CIs overlap, but because the model allows a single 
parameter  $\beta_{1,3}$ to be responsible for any difference in the upper asymptotes it becomes straightforward to detect. 

We next  note that  hemoglobin A1c\% is commonly used as a biomarker for diabetes mellitus,  with A1c\% $\geq 6.5\%$ indicating diabetes and  $5.7\% \geq$ A1c\% $< 6.5\%$ indicating prediabetes according
to guidelines of the American Diabetes Association (ADA)  \citep{malkaniAJM, 10.1001/jama.2021.12531}.  These thresholds are superimposed in the response curve in Figure \ref{figDiabetes2}. Interestingly, most of the rapid increase in the response curves occurs in the prediabetes range of  $5.7\% -  6.5\%$, while the upper asymptote is essentially reached just above the  $6.5\%$ threshold. Thus, the estimated model conforms very closely 
to the ADA guidelines. 

Finally, we note that the estimated model predicts a very high false positive rate for asthma positive subjects. That an association exists among hyperglycemia (and therefore elevated hemoglobin A1c\%)  and asthma 
has been consistently reported in the literature \citep{suissa2010, yang2020, diagnostics14171869, wen2025blood}.  This association is complex, and is described as ``bidirectional" in \cite{diagnostics14171869}. That the use of inhaled corticosteroids to control asthma and chronic obstructive pulmonary disease (COPD) may be associated with the  induction and worsening of diabetes has also been reported 
\citep{suissa2010, diagnostics14171869}. It seems possible, therefore, that this association may be present in subjects not diagnosed with diabetes \citep{suissa2010, yang2020} to a significant degree, leading to the higher 
false positive rate among asthma subjects  reported above (note that diagnoses in the MIDUS 3 study are self-reported).

\section{Discussion}   
  
In this article a collection of statistical methods extending the double-asymptotic binary response model  was proposed.  These models may have correlated responses, and can include additional covariates which potentially modify the components of the model, including the lower and upper asymptote.  This is achieved through what we define as the \emph{compound logistic regression model},  in which the mean response is a differential function of several logistic regression functions.  This permits a greater variety of models, while retaining the advantages of standard logistic regression.  The collection of methods is implemented in the \ttt{R}-package \ttt{CLmodel} (\url{https://github.com/ALMUDEVAR163/CLmodel/}). Additional model features are demonstrated in the vignette.

\bibliographystyle{rss}
\bibliography{cop}

 %\newpage
 
\end{document}